\documentclass[conference]{IEEEtran}

\usepackage{amsmath,amsfonts}
\usepackage{algorithmic}
\usepackage{graphicx}
\usepackage{textcomp}
\usepackage{xcolor}
\usepackage{cleveref}
\usepackage{enumitem}
\usepackage[acronym]{glossaries}
\usepackage{multirow}
\usepackage{xspace}
\usepackage{microtype}
\usepackage{pifont}
\usepackage{cite}
\usepackage[a4paper, total={184mm,239mm}]{geometry}
\usepackage{rotating}
\usepackage{pgfplots}
\usepackage{svg}
\usepgfplotslibrary{fillbetween}
\usepackage{multirow}
\usepackage{colortbl}
\usepackage[normalem]{ulem}
\useunder{\uline}{\ul}{}
\usepackage{lscape}
\usepackage{makecell}
\usepackage{threeparttable}

\crefname{figure}{Fig.}{Figs.}

\usepackage[free-standing-units,per-mode=repeated-symbol]{siunitx}
\usepackage[allow-number-unit-breaks]{siunitx}
\usetikzlibrary{scopes, calc, shapes, arrows, positioning, patterns}
\tikzset{>=latex}

\DeclareSIUnit\flop{FLOP}
\DeclareSIUnit\flops{FLOPS}
\DeclareSIUnit\gate{GE}
\DeclareSIUnit\op{OP}
\DeclareSIUnit\macu{MACU}
\DeclareSIUnit\ops{OPS}
\DeclareSIUnit\core{core}
\DeclareSIUnit\request{request}
\DeclareSIUnit\cycle{cycle}
\DeclareSIUnit\teraops{TOPS}
\DeclareSIUnit\ghz{GHz}
\DeclareSIUnit\mhz{MHz}
\DeclareSIUnit[number-unit-product = ]\percent{\%}

\makeatletter \newcommand{\AddSpaceIfAnonymous}{\@ifclasswith{acmart}{anonymous}{\vspace{10mm}}{}} \makeatother
\def\BibTeX{{\rm B\kern-.05em{\sc i\kern-.025em b}\kern-.08em
    T\kern-.1667em\lower.7ex\hbox{E}\kern-.125emX}}

\definecolor{s1dblu}{HTML}{706d94}
\definecolor{s1blu}{HTML}{6888a5}
\definecolor{s1gre}{HTML}{79b4a0}
\definecolor{s1lgre}{HTML}{a3c8a4}
\definecolor{s1greyel}{HTML}{d4daa1}
\definecolor{s1yel}{HTML}{ecd09c}
\definecolor{s1lora}{HTML}{e0a981}
\definecolor{s1ora}{HTML}{d37b6d}
\definecolor{s1red}{HTML}{b96570}
\definecolor{s1dred}{HTML}{8f5362}

\definecolor{cdblu}{rgb}{0.118,0.274,0.584}
\definecolor{cblu}{rgb}{0.321,0.561,0.678}
\definecolor{clblu}{rgb}{0.565,0.737,0.835}
\definecolor{cllblu}{rgb}{0.667,0.863,0.878}
\definecolor{clyel}{rgb}{1.000,0.902,0.718}
\definecolor{cyel}{rgb}{1.000,0.816,0.435}
\definecolor{clora}{rgb}{0.969, 0.667, 0.245}
\definecolor{cora}{rgb}{0.937,0.541,0.278}
\definecolor{cred}{rgb}{0.906,0.384,0.329}
\definecolor{clgay}{rgb}{0.6,0.6,0.6}

\definecolor{c3color0}{rgb}{8,29,89}
\definecolor{c3color1}{rgb}{33,49,140}
\definecolor{c3color2}{rgb}{34,84,163}
\definecolor{c3color3}{rgb}{30,128,184}
\definecolor{c3color4}{rgb}{48,165,194}
\definecolor{c3color5}{rgb}{91,191,192}
\definecolor{c3color6}{rgb}{149,213,184}
\definecolor{c3color7}{rgb}{205,235,179}
\definecolor{c3color8}{rgb}{238,248,180}
\definecolor{c3color9}{rgb}{254,254,215}

\begin{document}

\title{TCDM Burst Access: Breaking the Bandwidth Barrier in Shared-L1 RVV Clusters Beyond 1000 FPUs}


\ifx\blind\undefined
    \author{\IEEEauthorblockN{Diyou Shen\textsuperscript{*}}
    \IEEEauthorblockA{Integrated Systems Laboratory \\
    ETH Z\"urich \\
    Z\"urich, Switzerland \\
    dishen@iis.ee.ethz.ch}
    \and
    \IEEEauthorblockN{Yichao Zhang\textsuperscript{*}}
    \IEEEauthorblockA{Integrated Systems Laboratory \\
    ETH Z\"urich \\
    Z\"urich, Switzerland \\
    yiczhang@iis.ee.ethz.ch}
    \and
    \IEEEauthorblockN{Marco Bertuletti}
    \IEEEauthorblockA{Integrated Systems Laboratory \\
    ETH Z\"urich \\
    Z\"urich, Switzerland \\
    mbertuletti@iis.ee.ethz.ch}
    \and
    \IEEEauthorblockN{Luca Benini}
    \IEEEauthorblockA{Integrated Systems Laboratory \\
    ETH Z\"urich \\
    Z\"urich, Switzerland \\
    Universit\`a di Bologna \\
    Bologna, Italy \\
    lbenini@iis.ee.ethz.ch}
    }
\else
    \author{\centering{\textit{Authors omitted for blind review.}\vspace{2.5cm}}}
\fi

\maketitle

\begingroup
\renewcommand\thefootnote{\textsuperscript{*}} 
\footnotetext{\ifx\blind\undefined These two authors contributed equally to this work. \else Author information omitted for blind review \fi}
\endgroup

\newacronym{5G}{5G}{5th Generation}
\newacronym{6G}{6G}{6th Generation}
\newacronym{SDR}{SDR}{Software Defined Radio}
\newacronym{TTI}{TTI}{Transmission Time Interval}
\newacronym{RAN}{RAN}{Radio-Access-Networks}
\newacronym[longplural={Scratchpad Memories}]{SPM}{SPM}{Scratchpad Memory}
\newacronym{ACE}{ACE}{AXI Coherent Extensions}
\newacronym{AI}{AI}{Artificial Intelligence}
\newacronym{AMBA}{AMBA}{Advanced Microcontroller Bus Architecture}
\newacronym{AMX}{AMX}{Advanced Matrix Extension}
\newacronym{APB}{APB}{Advanced Peripheral Bus}
\newacronym{API}{API}{Application Programming Interface}
\newacronym{ASIC}{ASIC}{Application-Specific Integrated Circuit}
\newacronym{AVX}{AVX}{Advanced Vector Extension}
\newacronym{AXI}{AXI}{Advanced eXtensible Interface}
\newacronym{BLAS}{BLAS}{Basic Linear Algebra Subprograms}
\newacronym{CC}{CC}{Core Complex}
\newacronym{CHI}{CHI}{Coherent Hub Interface}
\newacronym{CMOS}{CMOS}{Complementary Metal-Oxide-Semiconductor}
\newacronym{CNN}{CNN}{Convolutional Neural Network}
\newacronym{CPU}{CPU}{Central Processing Unit}
\newacronym{CSR}{CSR}{Control and State Register}
\newacronym{CTS}{CTS}{Clock Tree Synthesis}
\newacronym{DLP}{DLP}{Data Level Parallelism}
\newacronym{DMA}{DMA}{Direct Memory Access}
\newacronym{DRAM}{DRAM}{Dynamic Random-Access Memory}
\newacronym{DSA}{DSA}{Domain-Specific Accelerator}
\newacronym{DSP}{DSP}{Digital Signal Processing}
\newacronym{DUT}{DUT}{Device Under Test}
\newacronym{DOTP}{DotP}{Dot Product}
\newacronym{ECL}{ECL}{Emitter-Coupled Logic}
\newacronym{FBB}{FBB}{Forward Body-Biasing}
\newacronym{FC}{FC}{Fully-Connected}
\newacronym{FFT}{FFT}{Fast Fourier Transform}
\newacronym{FDSOI}{FD-SOI}{Fully Depleted Silicon on Insulator}
\newacronym{FMA}{FMA}{Fused Multiply-Add}
\newacronym{FPGA}{FPGA}{Field-Programmable Gate Array}
\newacronym{FPU}{FPU}{Floating Point Unit}
\newacronym{FU}{FU}{Functional Unit}
\newacronym{FIFO}{FIFO}{first-in first-out}
\newacronym{GEMM}{GEMM}{General Matrix Multiply}
\newacronym{GPGPU}{GPGPU}{General-Purpose \acrlong{GPU}}
\newacronym{GPU}{GPU}{Graphics Processing Unit}
\newacronym{HDL}{HDL}{Hardware Description Language}
\newacronym{HERO}{HERO}{Heterogeneous Embedded Research Platform}
\newacronym{HPC}{HPC}{High-Performance Computing}
\newacronym{IoT}{IoT}{Internet of Things}
\newacronym{ILP}{ILP}{Instruction Level Parallelism}
\newacronym{IOT}{IoT}{Internet-of-Things}
\newacronym{IPC}{IPC}{Instructions Per Cycle}
\newacronym{IPU}{IPU}{Image Processing Unit}
\newacronym{ISA}{ISA}{Instruction Set Architecture}
\newacronym{LSU}{LSU}{Load/Store Unit}
\newacronym{LLM}{LLM}{Large Language Model}
\newacronym{LVT}{LVT}{low voltage threshold}
\newacronym{MATMUL}{MatMul}{Matrix Multiplication}
\newacronym{GE}{GE}{Gate Equivalents}
\newacronym{GF}{GF}{Grouping Factor}
\newacronym{MIMD}{MIMD}{multiple instruction, multiple data}
\newacronym{ML}{ML}{Machine Learning}
\newacronym{MMA}{MMA}{Matrix-Multiply Assist}
\newacronym{MME}{MME}{Matrix Multiplication Extension}
\newacronym{MMU}{MMU}{Memory Management Unit}
\newacronym{MRF}{MRF}{Matrix Register File}
\newacronym{MUL}{MUL}{multiplier}
\newacronym{MVL}{MVL}{maximum vector length}
\newacronym{NUMA}{NUMA}{Non-Uniform Memory Access}
\newacronym{NOC}{NoC}{Network-on-Chip}
\newacronym{MX}{MX}{Matrix eXtension}
\newacronym{PCIe}{PCIe}{Peripheral Component Interconnect Express}
\newacronym{PC}{PC}{Program Counter}
\newacronym{PE}{PE}{processing element}
\newacronym{PiM}{PiM}{Processing in memory}
\newacronym{PL}{PL}{Programmable Logic}
\newacronym{PMCA}{PMCA}{Programmable Manycore Accelerator}
\newacronym{PnM}{PnM}{Processing near memory}
\newacronym{PNR}{PnR}{Place-and-Route}
\newacronym{PSL}{PSL}{Power Service Layer}
\newacronym{PTE}{PTE}{page-table entry}
\newacronym{PTW}{PTW}{page-table walker}
\newacronym{PULP}{PULP}{Parallel Ultra Low Power}
\newacronym{QoS}{QoS}{quality of service}
\newacronym{RAW}{RAW}{read-after-write}
\newacronym{RBB}{RBB}{Reverse Body-Biasing}
\newacronym{ROB}{ROB}{Reorder Buffer}
\newacronym{RTL}{RTL}{Register Transfer Level}
\newacronym{RVT}{RVT}{Regular Voltage Threshold}
\newacronym{RVV}{RVV}{RISC-V Vector}
\newacronym{RoCC}{RoCC}{Rocket Custom Coprocessor Interface}
\newacronym{SCM}{SCM}{Storage Class Memory}
\newacronym{SIMD}{SIMD}{single instruction, multiple data}
\newacronym{SIMT}{SIMT}{single instruction, multiple thread}
\newacronym{SLDU}{SLDU}{Slide Unit}
\newacronym{SLVT}{SLVT}{super-low voltage threshold}
\newacronym{SM}{SM}{Streaming Multiprocessor}
\newacronym{SME}{SME}{Scalable Matrix Extension}
\newacronym{SOC}{SoC}{System-on-Chip}
\newacronym{SRAM}{SRAM}{Static Random-Access Memory}
\newacronym{SSE}{SSE}{Streaming SIMD Extension}
\newacronym{SVE}{SVE}{Scalable Vector Extension}
\newacronym{TCDM}{TCDM}{Tightly Coupled Data Memory}
\newacronym{TLP}{TLP}{Thread Level Parallelism}
\newacronym{TxnID}{TxnID}{Transaction ID}
\newacronym{VAC}{VAC}{Vector Access}
\newacronym{VC}{VC}{virtual channel}
\newacronym{VCONV}{VCONV}{Vector Conversion}
\newacronym{VEX}{VEX}{Vector Execute}
\newacronym{VFU}{VFU}{vector functional unit}
\newacronym{VID}{VID}{Vector Instruction Decode}
\newacronym{VIS}{VISSUE}{Vector Instruction Issue}
\newacronym{VLEN}{VLEN}{vector length}
\newacronym{VLIW}{VLIW}{Very Long Instruction Word}
\newacronym{VLOOP}{VLOOP}{Vector Loop}
\newacronym{VLR}{VLR}{vector length register}
\newacronym{VLSU}{VLSU}{Vector Load/Store Unit}
\newacronym{VNB}{VNB}{Von Neumann Bottleneck}
\newacronym{VRF}{VRF}{Vector Register File}
\newacronym{VPU}{VPU}{Vector Processing Unit}
\newacronym{VT}{VT}{vector thread}
\newacronym{WAR}{WAR}{write-after-read}
\newacronym{WAW}{WAW}{write-after-write}
\newacronym{DCT}{DCT}{discrete cosine transform}
\newacronym{TSV}{TSV}{through-silicon via}
\newacronym{3DIC}{3D-IC}{three-dimensional integrated circuit}
\newacronym{PPA}{PPA}{power, performance, and area}
\newacronym{F2F}{F2F}{face-to-face}
\newacronym{W2W}{W2W}{wafer-to-wafer}
\newacronym{IC}{IC}{integrated circuit}
\newacronym{C4}{C4}{controlled collapse chip connection}
\newacronym{FEOL}{FEOL}{front end of the line}
\newacronym{BEOL}{BEOL}{back end of the line}
\newacronym{PDP}{PDP}{power-delay product}
\newacronym{EDP}{EDP}{energy-delay product}
\newacronym{DRV}{DRV}{design rule violation}
\newacronym{DDR}{DDR}{double data rate}
\newacronym{SDRAM}{SDRAM}{synchronous dynamic random-access memory}
\newacronym{TPU}{TPU}{Tensor-Processing Unit}

\begin{abstract}
As computing demand and memory footprint of deep learning applications accelerate, clusters of cores sharing local (L1) multi-banked memory are widely used as key building blocks in large-scale architectures.
When the cluster's core count increases, a flat all-to-all interconnect between cores and L1 memory banks becomes a physical implementation bottleneck, and hierarchical network topologies are required.
However, hierarchical, multi-level intra-cluster networks are subject to internal contention which may lead to significant performance degradation, especially for SIMD or vector cores, as their memory access is bursty.
We present the TCDM Burst Access architecture, a software-transparent burst transaction support to improve bandwidth utilization in clusters with many vector cores tightly coupled to a multi-banked L1 data memory.
In our solution, a Burst Manager dispatches burst requests to L1 memory banks, multiple 32b words from burst responses are retired in parallel on channels with parametric data-width.
We validate our design on a \gls{RVV} many-core cluster, evaluating the benefits on different core counts.
With minimal logic area overhead (less than \SI{8}{\percent}), we improve the bandwidth of a 16-, a 256-, and a 1024-\gls{FPU} baseline clusters, without \gls{TCDM} Burst Access, by \SI{118}{\percent}, \SI{226}{\percent}, and \SI{77}{\percent} respectively.
Reaching up to \SI{80}{\percent} of the cores-memory peak bandwidth, our design demonstrates ultra-high bandwidth utilization and enables efficient performance scaling.
Implemented in \num{12}-nm FinFET technology node, compared to the serialized access baseline, our solution achieves up to \textbf{1.9x} energy efficiency and \textbf{2.76x} performance in real-world kernel benchmarkings.
\end{abstract}
\glsresetall


\section{Introduction}
\label{sec:introduction}
In the last ten years, the explosive growth of deep-learning workloads has driven the demand for parallel computing systems with large compute power and memory footprint: computing capacity requirements for training and inference of \gls{ML} models doubled every \num{6}-\num{9} months~\cite{Sevilla2022}, skyrocketing to quadrillion FLOPs; \glspl{LLM} require substantial memory capacity to handle hundreds of billions of parameters~\cite{llmsurvey2024}. 

Shared L1-memory clusters with programmable \glspl{PE} became a common architectural pattern to achieve both performance and energy efficiency: as the core count scales up, the memory access latency is kept low.
To increase the \gls{PE}-memory bandwidth, keeping the physical design modular and feasible, these architectures often incorporate a pipelined hierarchical interconnect, configuring a \gls{NUMA} system.
However, the limited number of ports to the shared interconnect nodes constrains the interconnection bandwidth, restricting the design scalability. 
Examples can be found in modern designs, where the size of shared-L1-memory clusters is limited to tens of \gls{PE}:
The TensTorrent architecture~\cite{Tenstorrent} packs only five \glspl{PE} with shared memory into a Tensix \textit{core warp}, forming the design's second hierarchy; In ET-SOC-1~\cite{ETSOC} by Esperanto, four 8-core \textit{Neighborhood} blocks shared only 4 \gls{SPM} banks through two hierarchical \gls{FC} crossbars; The Fujitsu A64RX~\cite{A64FX} and Kalray MPPA architecture~\cite{Kalray} have a shared memory across \num{13} and \num{16} cores, respectively.

Larger scale clusters are desirable to increase the tiling size of computations, thereby reducing data movement overheads, and increasing compute vs. memory transfer ratio in kernels like matrix multiplication (where this ratio grows as $N^3$/$N^2$)~\cite{terapool, mempool}.
The PULP-Platform's MemPool architecture~\cite{mempool} implements this concept by leveraging \num{256} \glspl{PE} tightly coupled to shared-L1, specifically \gls{TCDM}, where \gls{FC} crossbar ensures low-latency access between \glspl{PE} and banks.
Another key trend [9] is to increase the \glspl{PE} compute efficiency by leveraging vector \glspl{ISA}, which boosts the compute-fetch balance and increases the utilization of computing units.
However, vector load and stores need to access large chunks of consecutive addresses to keep the vector lanes busy. 
This translates into several simultaneous requests to the ports of the \gls{PE}-to-L1-memory interconnect.
As a result, the conflicts between vector PEs trying to concurrently access shared memory through the memory-to-\gls{PE} interconnect become critical and can bottleneck, severely limiting design scalability.

\begin{figure}[htbp]
    \centerline{
        \includegraphics[width=0.5\textwidth]{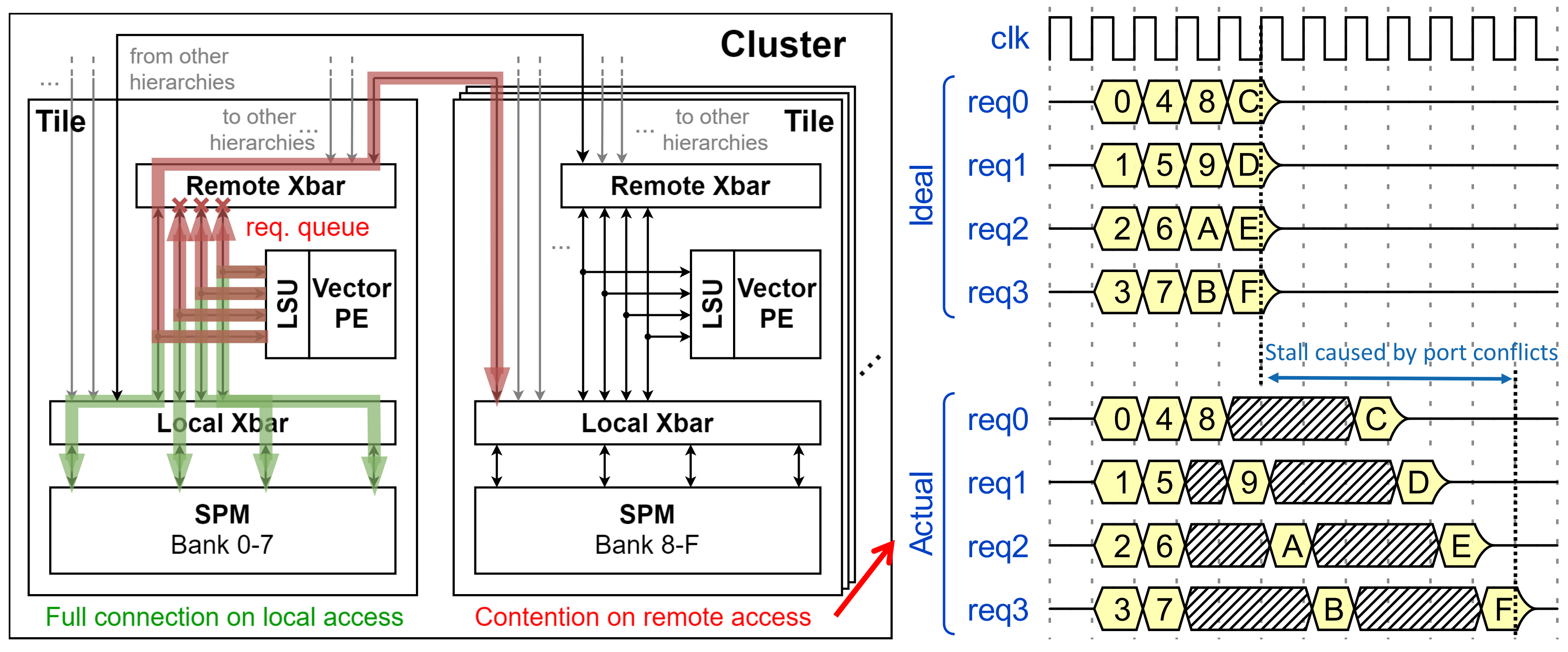}
    }
    \caption{Conflicts to shared interconnection resources reduce the interconnection bandwidth in vector many-core shared memory processors. The number on the request indicates its target bank.}
    \label{fig:port_confilct}
\end{figure}

\Cref{fig:port_confilct} illustrates the problem.
First note that the \gls{PE}-to-memory interconnect is hierarchical for physical scalability.
At the lowest level of the hierarchy, the tile, \glspl{PE} access a subset of the L1 shared memory (banks \num{0}-\num{7} in the example) with full bandwidth. 
However, accesses to banks in other tiles have to go through the remote crossbars with a number of remote ports smaller than the number of local ports (one to four in the example).
As a consequence, data fetching is serialized and fails to keep all \gls{LSU} ports busy, thereby underutilizing the full memory bandwidth of the processor.
Multiple approaches have been explored to mitigate this effect. 
Task scheduling techniques reduce memory access and alleviate interconnection pressure~\cite{task_schedule_2017}~\cite{task_schedule_2018}.
However, task-level optimization cannot fundamentally resolve the accessing conflict in the core-memory interconnection.
An architecture-aware data arrangement leverages the available interconnection ports better and reduces data movement overhead~\cite{data_arrange_2023}.
Nonetheless, this solution is typically tied to specific application algorithms, and adds complexity to software-managed data-allocation and transfer.
Topology-level optimization, such as 2D meshes typically implemented with \glspl{NOC}, can achieve high link bandwidth~\cite{FlooNoC_2023}. 
However, mesh-like \glspl{NOC} are unsuitable for interconnections between \glspl{PE} and L1 due to the additional latency from router hops, which substantially reduces the throughput available to the \glspl{PE} when traffic is not localized in tight sub-meshes.

In this paper, we propose the \gls{TCDM} Burst Access that breaks the interconnection bandwidth barrier caused by port competition in vector many-core clusters and helps the shared-L1 cluster to efficiently scale up beyond 1000 \glspl{FPU}.
We implement and validate our design approach across various scales of the MemPool-Spatz design~\cite{spatz}, an open-sourced, scalable many-core \gls{RVV} cluster.
Our results demonstrate a bandwidth improvement up to \SI{80}{\percent} of the theoretical design while maintaining minimal area impact and superior energy efficiency.
The main contributions of this paper are:
\begin{itemize}[leftmargin=*]
    \item
    The \gls{TCDM} Burst Access, a conflict-free mechanism on memory requests, implementing narrow-word (\num{32}-bit) burst accesses to L1 shared memory. A Burst Manager module designed to: (i) dispatch burst requests to a multi-banked scratchpad. (ii) merge the parallel memory responses into a single transaction, saturating the available interconnection bandwidth.
    \item 
    A physical-design aware \gls{FC} interconnect that maximizes the area-utilization of routing resources, by increasing data-width on the response channels only, to reduce serialization of burst responses.
    \item
    The validation of our design on a scalable \gls{RVV} cluster with different core-counts. Clusters with \num{16}/\num{256}/\num{1024} FPUs, obtain \SI{118}{\percent}, \SI{226}{\percent} and \SI{77}{\percent} bandwidth improvement respectively. Compared to the baseline, we achieve \SI{176}{\percent}, \SI{64}{\percent} and \SI{62}{\percent} performance improvement on real-world kernels: \gls{DOTP}, \gls{FFT}, and \gls{MATMUL}.
\end{itemize}
Implemented in GF12nm FinFET technology, our approach demonstrates less than \SI{8}{\percent} logic area overhead without introducing critical timing paths.
It improves up to \SI{90}{\percent} energy efficiency for memory-bound kernels. Our design is fully open-sourced\footnote{\ifx\blind\undefined https://github.com/pulp-platform/mempool \else Open-source information omitted for blind review. \fi}.

\section{Testbed Cluster and Peak-BW Analysis}
\label{sec:motivation}
To investigate the internal contention in hierarchical \glspl{PE}-memory networks, this section presents a bandwidth analysis based on a \gls{SIMD} many-core testbed cluster.
We analyze the theoretical peak interconnection bandwidth across various cluster scales, quantify the loss of bandwidth utilization, and outline our proposed solution.

\subsection{Testbed cluster architecture}
As discussed in~\cref{sec:introduction}, many-core vector clusters are susceptible to interconnect contention due to \gls{SIMD} load and store operations accessing consecutive addresses.
This leads to conflicts at the same ports of the hierarchical interconnection.
We select \textit{MemPool-Spatz}~\cite{spatz} as our testbed architecture, an open-sourced, scalable, \gls{RVV} many-core shared-L1 cluster based on the \verb#Zve32f# \gls{ISA}.
The architecture's \glspl{PE} are \gls{CC}, where one \textit{Snitch} scalar core is responsible for executing scalar instructions and forwarding vector instructions to a floating point \textit{Spatz} vector core.
We follow the naming convention \textit{MP\textsubscript{N}Spatz\textsubscript{K}} to represent different design scales, where \textit{N} indicates the number of \glspl{CC} and \textit{K} represents the number of vector \glspl{FPU} per \textit{Spatz} vector core.
The total number of \glspl{FPU} is given by $\textit{N} \times \textit{K}$.

In the hierarchical multi-level interconnection design of \textit{MP\textsubscript{N}Spatz\textsubscript{K}}, all \glspl{PE} have shared \gls{NUMA} to $N\times4$ fully interleaved \SI{1}{\kibi\byte} banks of \gls{SPM}.
In this paper, we focus on the two 16 and 256-\gls{FPU} most energy-efficient configurations demonstrated by~\cite{spatz}.
We further scale it up to a 1024-\gls{FPU} configuration, incorporating a hierarchy configuration inspired by~\cite{terapool}, as follows:
\begin{enumerate}
    \item \textit{MP\textsubscript{4}Spatz\textsubscript{4}}: a \num{16}-FPU vector cluster with a maximum \gls{VLEN} of \SI{256}{bits}. The design is built with one hierarchy, the \textit{Tile}, consisting of \num{4} \glspl{CC} and \num{16} \gls{SPM} banks with \num{1}-cycle round-trip accessing latency. Each \textit{Tile} has four hierarchical interconnection ports, accessing other \textit{Tiles} with \num{3} cycles round-trip latency.
    \item \textit{MP\textsubscript{64}Spatz\textsubscript{4}}: a \num{256}-FPU vector cluster with a maximum \gls{VLEN} of \SI{256}{bits}. The design is built in two hierarchies. The \textit{Tile} hierarchy consists of \num{4} \glspl{CC} and \num{16} \gls{SPM} banks with \num{1}-cycle round-trip accessing latency. Four \textit{Group} hierarchy blocks, contain \num{16} \textit{Tiles} each. Each \textit{Tile} and \textit{Group} has four hierarchical interconnection ports accessing other \textit{Tiles} with \num{3} cycles, and other \textit{Groups} with \num{5} cycles round-trip latency.
    \item \textit{MP\textsubscript{128}Spatz\textsubscript{8}}: a \num{1024}-FPU vector cluster with a maximum \gls{VLEN} of \SI{512}{bits}. The design is built in three hierarchies. The \textit{Tile} hierarchy consists of \num{8} \glspl{CC} and \num{32} \gls{SPM} banks with \num{1}-cycle round-trip accessing latency, followed by four \textit{SubGroup} hierarchy blocks with \num{8} \textit{Tiles} each. Four \textit{SubGroups} form the \textit{Group} hierarchy. Each \textit{Tile} has seven hierarchical interconnection ports: one port accesses other \textit{Tiles} within the same \textit{SubGroup} with \num{3} cycles latency; three ports access other three \textit{Subgroups} within the same \textit{Group} with \num{5} cycles latency; and three ports access remote \textit{Groups} with \num{9} cycles round-trip latency.
\end{enumerate}

\subsection{Interconnect bandwidth analysis}
\label{sec:model}
\begin{figure}[htbp]
\centerline{
    \includegraphics[width=0.5\textwidth]{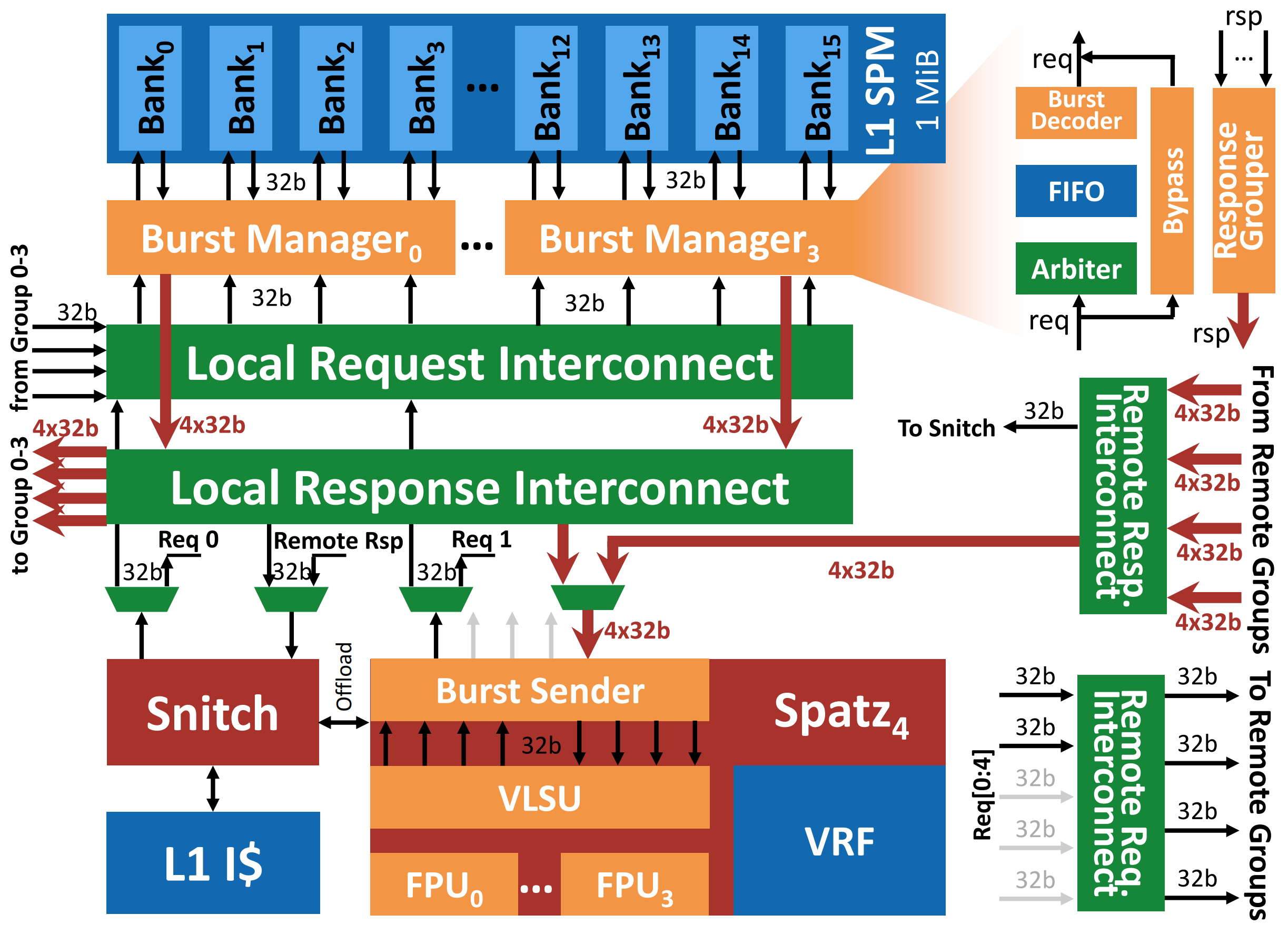}
}
\caption{MP\textsubscript{64}Spatz\textsubscript{4}'s Tile level architectural schematic with \gls{TCDM} burst and \textit{GF4}. The increased data-width response channels are marked in red.}
\label{block_diagram}
\end{figure}

The \gls{VLSU} manages the memory accesses of the vector core, with the number of request and response ports matching the number of $K$ \glspl{FPU} in the Spatz\textsubscript{K} design as shown in~\Cref{block_diagram}.
The \gls{VLSU} splits a vector memory request into multiple \SI{32}{b} data requests and distributes them across the available \gls{VLSU} ports.
The theoretical \gls{VLSU} peak bandwidth can be defined as the bandwidth achieved when all requests sent through the \gls{VLSU} are routed by an all-to-all fully connected crossbar without any contention:
\begin{equation}
    BW_{vlsuPeak} = BW_{Spatz_K} = K \times 4~Bytes/cyc
\end{equation}

In hierarchical \gls{FC} crossbars-based testbed clusters, memory accesses can be categorized as  \textit{local-Tile} or \textit{remote-Hierarchy}, depending on the requested target address.
In \textit{local-Tile} accesses, memory requests from a \gls{VLSU} target the \gls{SPM} banks within its own \textit{Tile}. 
This achieves full local-interconnection bandwidth if no bank conflicts are encountered, benefiting from a \gls{FC} crossbar that does not require arbitration.
In contrast, \textit{remote-Hierarchy} accesses encounter conflicts when the parallel requests from a \gls{VLSU} target L1 address portions that are allocated to the same shared interconnection port.
This results in decreased bandwidth utilization as parallel requests must be arbitrated and serialized, as previously illustrated in~\Cref{fig:port_confilct}.
The estimated bandwidth is as follows:
\begin{align}
    &{BW}_{locTile} = BW_{vlsuPeak} \\
    &{BW}_{rmtHier} = BW_{serialized} = 4~Bytes/cyc
    \label{bw_r}
\end{align}

Assuming each vector request targets random and uniformly distributed destination banks, with $N_{PE}$ representing the total number of vector cores, the probabilities of targeting \textit{local-Tile} and \textit{remote-Hierarchy} accesses are denoted as $p_l$ and $p_r$, respectively, as shown in~\cref{p_access}.
Additionally, the random accessing average bandwidth is presented in~\cref{bw}.
\begin{equation}
    p_l = \frac{1}{N_{PE}},\ \ p_r=1-p_{l}= \frac{N_{PE}-1}{N_{PE}}
    \label{p_access}
\end{equation}

\begin{equation}
    BW_{hierAvg} = \mathbb{E}[{BW}] = p_l \cdot {BW}_{locTile} + p_r \cdot {BW}_{rmtHier}
    \label{bw}
\end{equation}

We calculate the theoretical \gls{VLSU} peak bandwidth ($BW_{vlsuPeak}$) and the hierarchical interconnection average bandwidth with random accessing ($BW_{hierAvg}$) for all three scaled testbed clusters, and summarize the results in the first two rows of~\Cref{tab_BW_calc}.
The results demonstrate significantly lower bandwidth in the multi-level hierarchical \gls{FC} crossbar compared with the peak bandwidth that \gls{VLSU} interfaces could support.
In the MP\textsubscript{128}Spatz\textsubscript{8}, the local-Tile bandwidth increases, scaling with the number of \glspl{CC}.
As a result, the hierarchical interconnection average bandwidth of the baseline MP\textsubscript{128}Spatz\textsubscript{8} testbed cluster slightly improves, but the bandwidth utilization (\SI{11.75}{\percent}) reduces, due to the increased \gls{VLSU} peak bandwidth.
Thus, finding a solution to mitigate the hierarchical interconnection conflicts is crucial for maintaining performance scalability in large-scale shared-memory vector cluster designs.
In the following subsection, we introduce our proposed solution to address this challenge effectively.

\begin{table}[ht]
\caption{Calculated Memory Bandwidth: Comparison Across Cluster Sizes and Configurations.}
\label{tab_BW_calc}
\centering
\begin{tabular}{llrrr}
\hline
                                  &       & \multicolumn{1}{l}{\textbf{MP\textsubscript{4}Spatz\textsubscript{4}}} & \multicolumn{1}{l}{\textbf{MP\textsubscript{64}Spatz\textsubscript{4}}} & \multicolumn{1}{l}{\textbf{MP\textsubscript{128}Spatz\textsubscript{8}}} \\
\textbf{Peak}                    & BW {[}B/cyc{]}    & 16.00                                       & 16.00                                        & 32.00                                          \\ \hline
\multirow{2}{*}{\textbf{Baseline}} & BW {[}B/cyc{]}    & 7.00                                        & 4.18                                         & 4.22                                           \\
                                  & Utilization        & 37.50\%                                     & 21.38\%                                      & 11.75\%                                        \\ \hline
\multirow{4}{*}{\textbf{2xRsp}}   & BW {[}B/cyc{]}    & 10.00                                       & 8.13                                         & 8.19                                           \\
                                  & Utilization        & 62.50\%                                     & 50.78\%                                      & 25.59\%                                        \\
                                  & Improvement       & +42.86\%                                    & +94.38\%                                     & +94.02\%                                        \\ \hline
\multirow{4}{*}{\textbf{4xRsp}}   & BW {[}B/cyc{]}    & 16.00                                       & 16.00                                        & 16.13                                          \\
                                  & Utilization        & 100.00\%                                    & 100.00\%                                     & 50.39\%                                        \\
                                  & Improvement       & +128.57\%                                   & +282.78\%                                    & +282.11\%                                       \\ \hline
\end{tabular}
\end{table}

\subsection{Burst Access for TCDM conflicts reduction}
We propose the \textit{\gls{TCDM} Burst Access} as a solution to break the interconnection bandwidth barrier due to port competition, specifically enhancing the load request and the memory-response channels in multi-level hierarchical \gls{FC} crossbar.
We focus on loads because the latency of store operations is hidden by the synchronization time required to solve inter-core data dependencies in \gls{SPM}-based parallel clusters. Stores are consequently non-critical for the cluster performance.
Our solution focuses on loads and consists of two key contributions:

\subsubsection{Burst narrow requests}
To resolve the port contentions on hierarchical interconnect,  a widely used approach is to reduce the number of memory requests by employing a \textit{burst access} mechanism.
In this mechanism, multiple narrow memory requests (\SI{32}{b}) are combined into a single transfer with the \textit{burst length} information, which specifies the number of consecutive element words to be requested.
This is particularly advantageous for vector requests, as their consecutive address patterns can straightforwardly be mapped to a burst format by specifying a start address and burst length.

\subsubsection{Increased response data-width}
Upon receiving a burst request, the \gls{SPM} banks process requests simultaneously and generate data responses in parallel.
This scenario introduces port contentions on the memory-response channel, leading us to the second aspect of our solution:
the parallel response data can be merged in the memory-response channel with an expanded data field, thereby reducing the number of individual transfers sent across the interconnection.
However, the routing complexity in the \gls{FC} crossbar design linearly increases with the width of the data field.
To maintain physical feasibility in differently scaled clusters, the data width extension should remain hardware-configurable, allowing for flexible adjustments to ensure high routing resource utilization.
The \gls{GF} describes the multiplier used to extend the data width on the response channel. 
In this paper, we explore the bandwidth improvements associated with doubling (\textit{GF2}) and quadrupling (\textit{GF4}) the response channel data fields.
The results from the analytical model outlined in~\cref{sec:model} are presented in~\Cref{tab_BW_calc}.
According to our model, full bandwidth utilization becomes achievable when the width factor of the response data field equals the number of \gls{VLSU} ports.
Our proposed solution improves bandwidth in multi-level hierarchical interconnection designs and enhances bandwidth utilization during further scaling up.

\section{Architecture}
This section presents the key architectural components designed to support \gls{TCDM} Burst Access.
We implement \gls{TCDM} Burst Access in the testbed clusters' \textit{Tiles}.
A \textit{Tile} diagram with \textit{GF4} implementation on MP\textsubscript{64}Spatz\textsubscript{4} is shown as an example in~\Cref{block_diagram}.

\subsection{Burst Sender}
The Burst Sender is attached to the \gls{VLSU} ports in the Spatz processor.
When detecting a \texttt{VLE} instruction, it combines the \textit{K} parallel requests at \gls{VLSU} ports into a single burst with a burst length of \textit{K} words.
In the MemPool-Spatz testbed, the orders between memory requests and responses are guaranteed using \glspl{ROB} at \gls{VLSU} ports.
These \glspl{ROB} are also used for latency tolerance by enabling multiple outstanding transactions.
Since each burst request contains multiple narrow requests, the depth of the \gls{ROB} needs to be increased to maintain the same level of outstanding transaction support, which is doubled in our testbed clusters, as an example.

\subsection{Burst Manager}
We design a Burst Manager module that serves as a burst format adapter.
It efficiently splits or combines \SI{32}{b} narrow requests and responses, adapting them to \gls{SPM} banks without complicating the memory module design.
Further details are:
i)  On the request channel, the Burst Manager receives burst requests, converts them into parallel \SI{32}{b} memory requests, and forwards them to the \gls{SPM} banks.
If multiple bursts arrive simultaneously, an arbitrator and a small \gls{FIFO} buffer are used to hold the following burst requests.
ii) On the response channel, the Burst Manager leverages the widened response data width, configured through an elaboration parameter (\textit{GF}).
It merges the parallel response data into a single transfer and forwards it through the widened data field.
This block is needed for every \textit{GF} number of \gls{SPM} banks to handle the burst requests in parallel.

We implemented our solution with modular designs to minimize the changes in the original testbed clusters.
In both MP\textsubscript{4}Spatz\textsubscript{4} and MP\textsubscript{64}Spatz\textsubscript{4}, a \textit{GF4} design is implemented for maximizing the bandwidth.
A \textit{GF2} design is used in MP\textsubscript{128}Spatz\textsubscript{8} considering the increased routing congestion in scaling.
In the next section, we will evaluate the performance of our design as implemented on these testbeds.

\begin{figure*}
    \centering
    \begin{minipage}{0.32\textwidth}
        \centering
        \includegraphics[width=\linewidth]{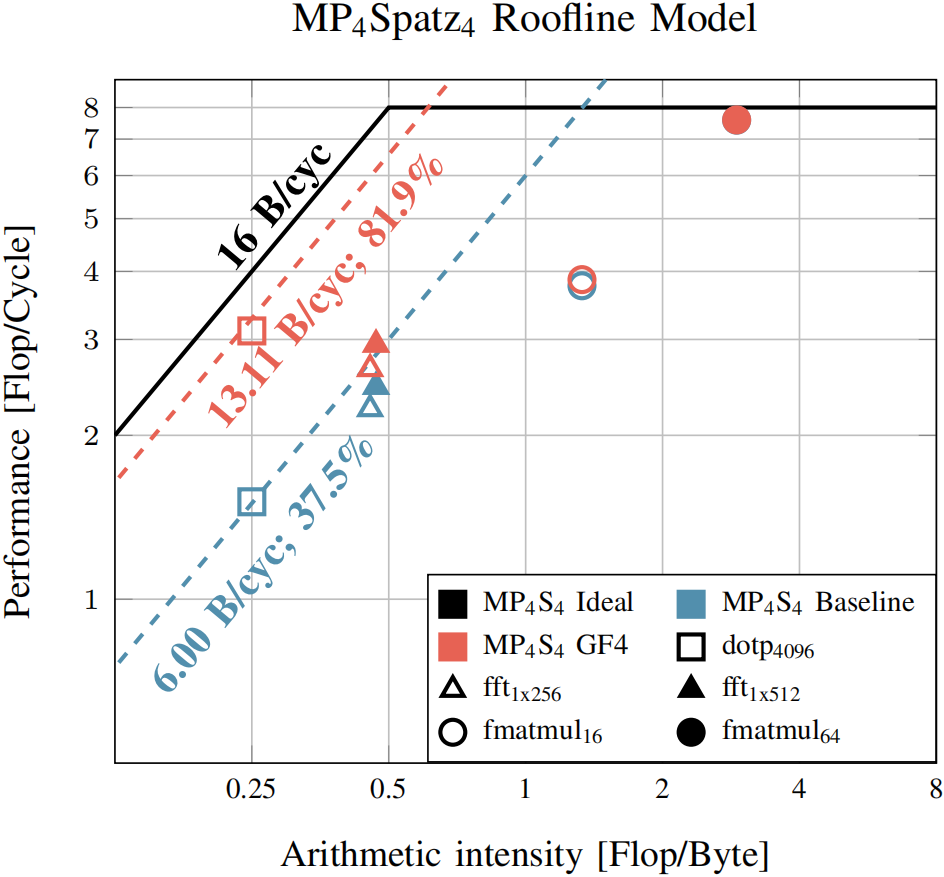}
    \end{minipage}
    \hfill
    \begin{minipage}{0.32\textwidth}
        \centering
        \includegraphics[width=\linewidth]{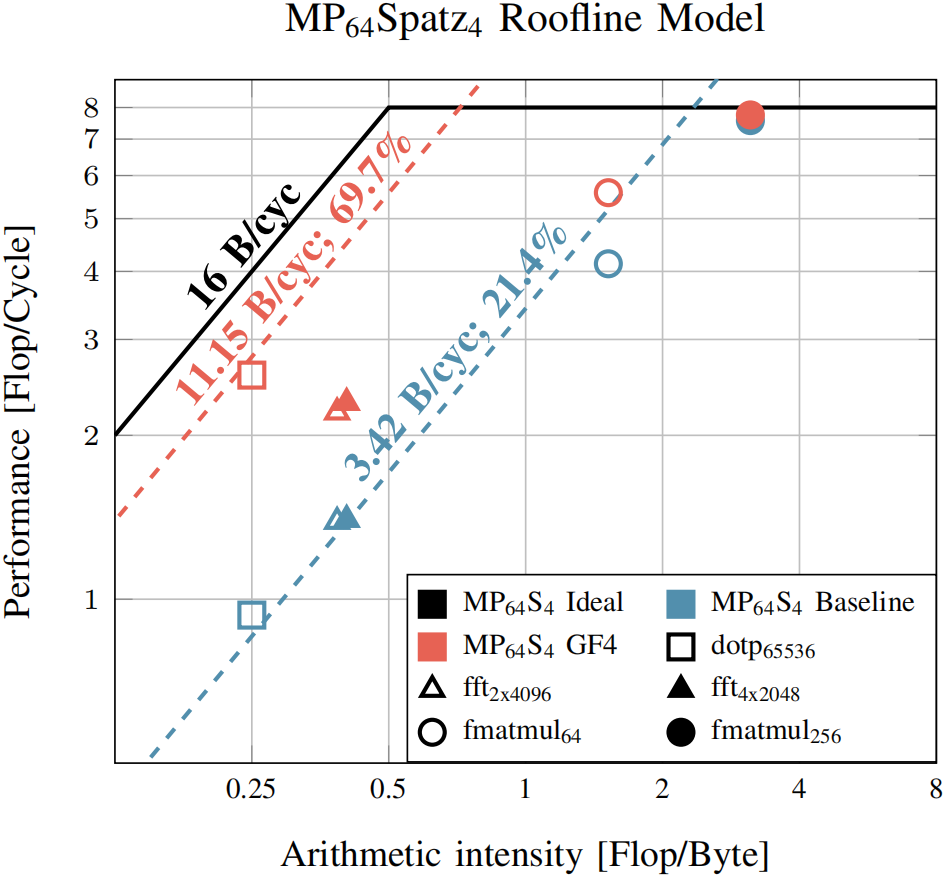}
    \end{minipage}
    \hfill
    \begin{minipage}{0.32\textwidth}
        \centering
        \includegraphics[width=\linewidth]{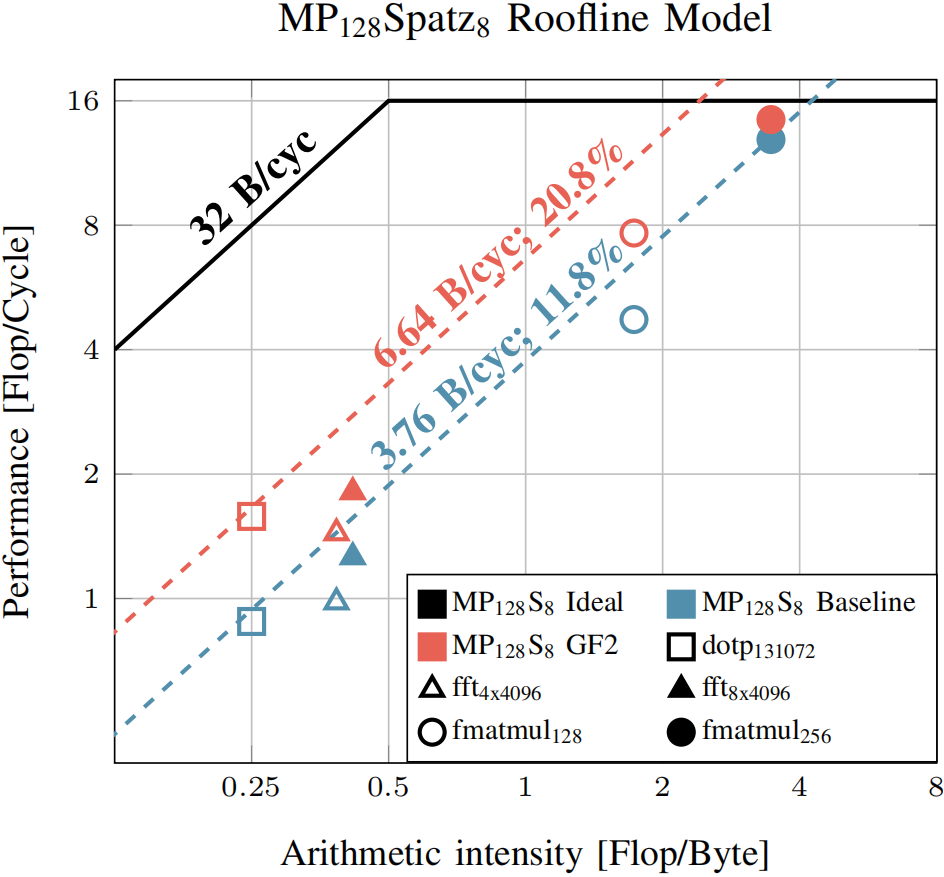}
    \end{minipage}
    \caption{The roofline plots on original and burst-enabled configurations on MP\textsubscript{4}Spatz\textsubscript{4} (left), MP\textsubscript{64}Spatz\textsubscript{4} (middle), and MP\textsubscript{128}Spatz\textsubscript{8} (right). The hierarchical average bandwidth is shown in dashed lines in the graph, the ideal no-contention bandwidth and maximum achievable performance are in solid black lines.}
    \label{fig:roofline}
\end{figure*}

\section{Performance Analysis}
The roofline model is widely used to analyze an architecture's performance with respect to the memory bandwidth~\cite{roofline}.
We present the roofline models of our designs on the testbed clusters in~\Cref{fig:roofline}.
We determined the ideal no-contention bandwidth per core using the theoretical \gls{VLSU} peak bandwidth, and the maximum achievable performance with the theoretical maximum throughput of the \glspl{FPU} in a \gls{PE}.

In our analysis, we benchmark different real-world kernels with distinct arithmetic intensities to demonstrate the effectiveness of \gls{TCDM} Burst Access mechanism in improving bandwidth utilization and performance:
\begin{enumerate}
    \item \textit{\gls{DOTP}}: Multiplication between two $n$-element vectors, with an arithmetic intensity of \SI{0.25}{FLOPs/byte}.
    \item \textit{\gls{FFT}}: Multi-core implementation of the Cooley-Tukey Radix-2 \gls{FFT} algorithm, running $k$ instances of $n$-point \glspl{FFT} in parallel across all cores on complex single-precision floating-point samples.
    Depending on both the problem size of \gls{FFT} and the number of cores involved, the arithmetic intensity ranges between \SI{0.3}{FLOPs/byte} and \SI{0.5}{FLOPs/byte}.
    \item \textit{\gls{MATMUL}}: Matrix multiplication on two $n \times n$ single-precision floating-point matrices.
    The arithmetic intensity varies depending on the problem sizes of the matrices.
    We evaluate the performance on two sizes of \gls{MATMUL} kernels on each hardware configuration, with arithmetic intensity of at \SI{1.5}{FLOPs/byte} and \SI{3.5}{FLOPs/byte}, respectively.
\end{enumerate}

Additionally, we simulate and present the bandwidth analysis model in~\cref{sec:model}, by using a test kernel with vector loads targetting random addresses, showing as the dashed line in~\Cref{fig:roofline}.
All kernels and tests follow a fork-join programming model, with all data preloaded into the testbed's L1 memory.

The roofline plots, shown in~\Cref{fig:roofline}, compare the testbed clusters with and without \gls{TCDM} Burst Access implementation.
Our \textit{GF4} design improves the hierarchical average bandwidth by \SI{118}{\percent} and \SI{226}{\percent}, achieving average bandwidth utilization of \SI{82}{\percent} and \SI{70}{\percent}, in MP\textsubscript{16}Spatz\textsubscript{4} and MP\textsubscript{64}Spatz\textsubscript{4} cluster, respectively.
In \gls{DOTP} kernel, the \textit{GF4} design shows a \SI{106}{\percent} and \SI{176}{\percent} performance improvement compared to the baseline testbed in MP\textsubscript{16}Spatz\textsubscript{4} and MP\textsubscript{64}Spatz\textsubscript{4}, respectively, closely matching the improvements in bandwidth.
Smaller gains of \SI{41}{\percent} and \SI{64}{\percent} are observed in \gls{FFT} kernel for MP\textsubscript{16}Spatz\textsubscript{4} and MP\textsubscript{64}Spatz\textsubscript{4}, due to the unavoidable inter-core synchronization inherent in the multi-core \gls{FFT} algorithm.
The performance of compute-bound \gls{MATMUL} kernels do not differ between the baseline testbed and our \textit{GF4} design in  MP\textsubscript{16}Spatz\textsubscript{4} cluster.
When working on the smaller matrix sizes in a \gls{MATMUL} kernel, the ratio of data transfer to computation becomes significant, causing the performance to be limited by the memory bandwidth.
In this scenario, such as the $64\times64\times64$ \gls{MATMUL} in MP\textsubscript{64}Spatz\textsubscript{4} cluster, a notable performance improvement of \SI{35}{\percent} is observed by implementing \textit{GF4} design.

The MP\textsubscript{128}Spatz\textsubscript{8} shows a higher hierarchical average bandwidth compared to the MP\textsubscript{64}Spatz\textsubscript{4} testbed cluster, consistent with the estimation in~\cref{sec:model}.
By implementing the \gls{TCDM} Burst Access with \textit{GF2} configuration, the hierarchical average bandwidth is improved by \SI{90}{\percent}, reaching the utilization of \SI{20.8}{\percent}.
The testbed with \textit{GF2} shows performance improvements of \SI{80}{\percent} and \SI{47}{\percent} in \gls{DOTP} and \gls{FFT} kernels compared to the baseline MP\textsubscript{128}Spatz\textsubscript{8} cluster.
The larger cluster scale requires a higher problem size of \gls{MATMUL} kernel to remain in the compute-bound region.
Because of this, a $128\times 128\times 128$ \gls{MATMUL} kernel achieves \SI{62}{\percent} performance improvement, a $256\times 256\times 256$ \gls{MATMUL} kernel moves into the compute-bound region, obtaining \SI{12}{\percent} improvement, and over \SI{90}{\percent} \gls{FPU} utilization.

\begin{table*}
    \caption{The Summary of Kernel Performance and Energy Efficiency.}
    \label{tab_kernel}
    \centering
\begin{threeparttable}
\begin{tabular}{llrrrrrrrr}
\Xhline{1.5pt}
\multicolumn{1}{c}{\textbf{Config}} &
  \multicolumn{1}{c}{\textbf{Kernel}} &
  \multicolumn{1}{c}{\textbf{Kernel Size}} &
  \textbf{\begin{tabular}[c]{@{}r@{}}Arithmetric\\ Intensity\\ {[}FLOP/B{]}\end{tabular}} &
  \textbf{\begin{tabular}[c]{@{}r@{}}FPU\\ Utilization\end{tabular}} &
  \textbf{\begin{tabular}[c]{@{}r@{}}Performance\\ @ss\_freq\\ {[}GFLOPS{]}\end{tabular}} &
  \textbf{\begin{tabular}[c]{@{}r@{}}Performance\\ @tt\_freq\\ {[}GFLOPS{]}\end{tabular}} &
  \textbf{\begin{tabular}[c]{@{}r@{}}Power\\ @tt\_freq\\ {[}W{]}\end{tabular}} &
  \textbf{\begin{tabular}[c]{@{}r@{}}En. Efficiency\\ @tt\_freq\\ {[}GFLOPS/W{]}\end{tabular}} &
  \textbf{\begin{tabular}[c]{@{}r@{}}En. Efficiency\\ Comparsion\end{tabular}} \\ \Xhline{1.5pt}
\multicolumn{10}{c}{\textbf{MP\textsubscript{4}Spatz\textsubscript{4} Cluster}\tnote{1}} \\ \hline
Baseline & dotp   & 4096        & 0.25 & 18.88\% & 4.65    & 5.50    & 0.09 & 63.12  & -        \\
Baseline & fft    & 1x512       & 0.47 & 30.71\% & 7.57    & 8.94    & 0.09 & 95.14  & -        \\
Baseline & matmul & 16x16x16    & 1.33 & 47.06\% & 11.60   & 13.70   & 0.12 & 118.70 & -        \\
Baseline & matmul & 64x64x64    & 2.91 & 94.97\% & 23.40   & 27.66   & 0.13 & 218.69 & -        \\ \hline
GF4      & dotp   & 4096        & 0.25 & 38.91\% & 9.59    & 11.33   & 0.12 & 91.82  & \textcolor{s1gre}{+45.47\%} \\
GF4      & fft    & 1x512       & 0.47 & 42.72\% & 10.53   & 12.44   & 0.13 & 96.58  & \textcolor{s1gre}{+1.52\%}  \\
GF4      & matmul & 16x16x16    & 1.33 & 48.30\% & 11.90   & 14.06   & 0.12 & 113.28 & \textcolor{s1dred}{-4.57\%}  \\
GF4      & matmul & 64x64x64    & 2.91 & 94.95\% & 23.40   & 27.65   & 0.13 & 206.82 & \textcolor{s1dred}{-5.43\%}  \\\Xhline{1.5pt}
\multicolumn{10}{c}{\textbf{MP\textsubscript{64}Spatz\textsubscript{4} Cluster}\tnote{1}}\\ \hline
Baseline & dotp   & 65536       & 0.25 & 12.06\% & 47.55   & 56.19   & 1.32 & 42.70  & -        \\
Baseline & fft    & 4x2048      & 0.37 & 17.51\% & 69.03   & 81.58   & 1.30 & 62.95  & -        \\
Baseline & matmul & 64x64x64    & 1.52 & 51.64\% & 203.59  & 240.60  & 1.45 & 166.05 & -        \\
Baseline & matmul & 256x256x256 & 3.12 & 94.58\% & 372.87  & 440.67  & 1.77 & 248.40 & -        \\ \hline
GF4      & dotp   & 65536       & 0.25 & 33.29\% & 131.24  & 155.10  & 1.91 & 81.12  & \textcolor{s1gre}{+89.99\%} \\
GF4      & fft    & 4x2048      & 0.37 & 28.70\% & 113.15  & 133.72  & 1.76 & 75.80  & \textcolor{s1gre}{+20.42\%} \\
GF4      & matmul & 64x64x64    & 1.52 & 69.75\% & 274.98  & 324.98  & 1.84 & 176.62 & \textcolor{s1gre}{+6.37\%}  \\
GF4      & matmul & 256x256x256 & 3.12 & 96.93\% & 382.14  & 451.62  & 1.97 & 229.01 & \textcolor{s1dred}{-7.81\%}  \\\Xhline{1.5pt}
\multicolumn{10}{c}{\textbf{MP\textsubscript{128}Spatz\textsubscript{8} Cluster}\tnote{2}}\\ \hline
Baseline & dotp   & 131072      & 0.25 & 5.49\%  & 71.28   & 98.38   & 4.24 & 23.20  & -        \\
Baseline & fft    & 4096x8      & 0.42 & 7.87\%  & 102.19  & 141.03  & 4.03 & 34.98  & -        \\
Baseline & matmul & 128x128x128 & 1.73 & 29.56\% & 383.82  & 529.72  & 7.30 & 72.52  & -        \\
Baseline & matmul & 256x256x256 & 3.46 & 80.57\% & 1046.15 & 1443.81 & 7.78 & 185.68 & -        \\ \hline
GF2      & dotp   & 131072      & 0.25 & 9.85\%  & 127.90  & 176.51  & 5.41 & 32.64  & \textcolor{s1gre}{+40.67\%} \\
GF2      & fft    & 4096x8      & 0.42 & 11.32\% & 146.98  & 202.85  & 4.62 & 43.87  & \textcolor{s1gre}{+25.42\%} \\
GF2      & matmul & 128x128x128 & 1.73 & 47.86\% & 621.43  & 857.65  & 8.14 & 105.40 & \textcolor{s1gre}{+45.34\%} \\
GF2      & matmul & 256x256x256 & 3.46 & 90.09\% & 1169.76 & 1614.41 & 8.91 & 181.15 & \textcolor{s1dred}{-2.44\%}  \\\Xhline{1.5pt}
\end{tabular}
\begin{tablenotes}
\item[1]{In MP\textsubscript{4}Spatz\textsubscript{4} and MP\textsubscript{64}Spatz\textsubscript{4}, ss\_freq = \SI{770}{MHz}, tt\_freq = \SI{910}{MHz}}
\item[2]{In MP\textsubscript{128}Spatz\textsubscript{8}, ss\_freq = \SI{634}{MHz}, tt\_freq = \SI{875}{MHz}}
\end{tablenotes}
\end{threeparttable}

\end{table*}

\section{Physical Implementation}
In this section, we analyze the \gls{PPA} of our design \gls{PNR} implementation.
\begin{figure}[ht]
    \centering
    \includegraphics[width=\linewidth]{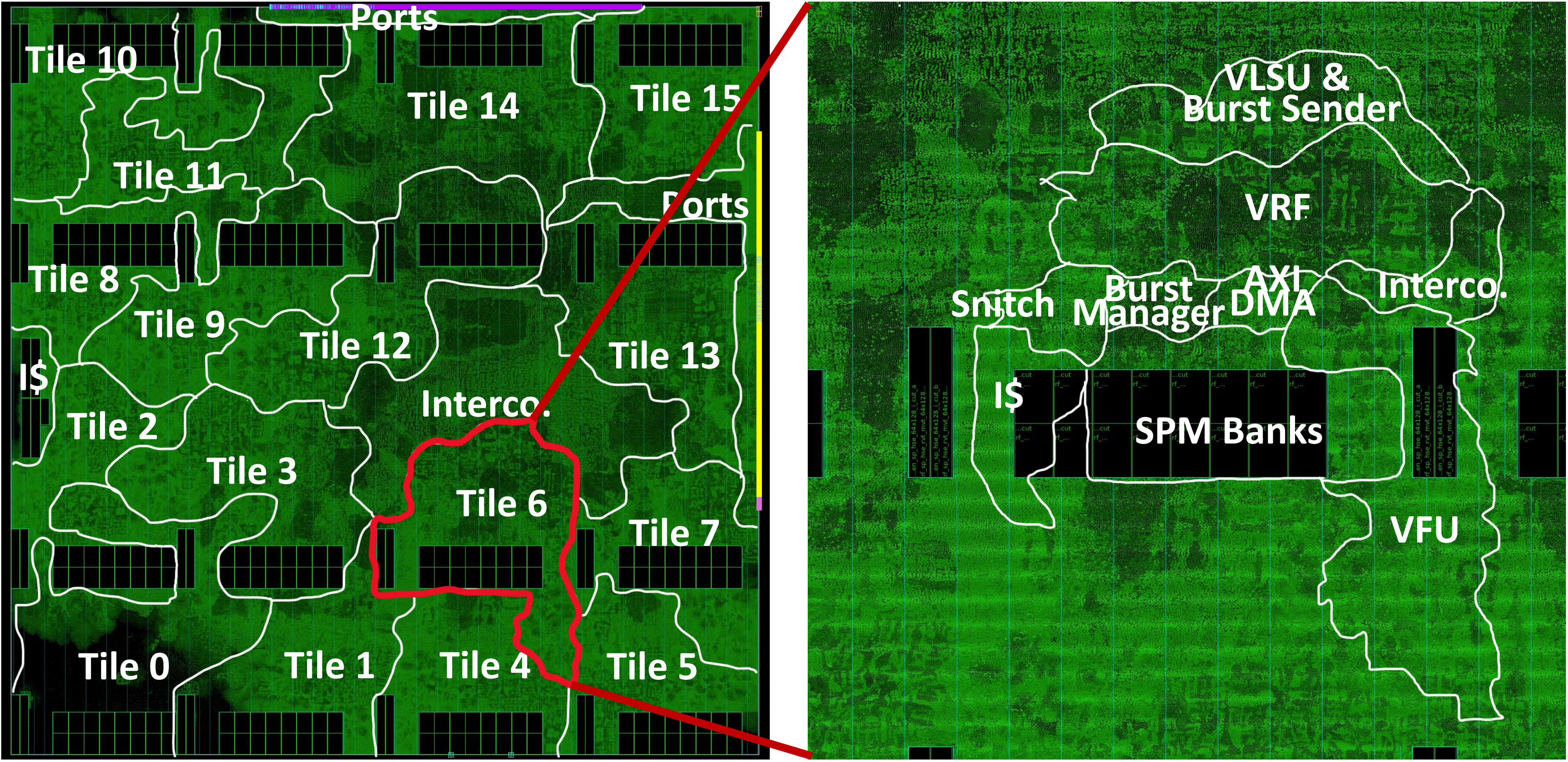}
    \caption{Placed-and-routed layout annotated Group- and Tile-level view of \textit{GF4} design on MP\textsubscript{64}Spatz\textsubscript{4} cluster}
    \label{fig:layout}
\end{figure}

We synthesize and \gls{PNR} (Synopsys Fusion Compiler 2022.03) the testbed clusters with \gls{TCDM} Burst Access mechanism in GlobalFoundries' 12nm LP-PLUS FinFET technology.
Both MP\textsubscript{16}Spatz\textsubscript{4} and MP\textsubscript{64}Spatz\textsubscript{4} are targeted to run at \SI{770}{MHz}, while MP\textsubscript{128}Spatz\textsubscript{8} targets \SI{634}{MHz} under worst-case conditions (SS/\SI{0.72}{V}/\SI{125}{\textsuperscript{$\circ$}C}), with no frequency degradation compared to the original testbed.
Power estimations are obtained using Synopsys PrimeTime 2022.03 under nominal conditions (TT/\SI{0.80}{V}/\SI{25}{\textsuperscript{$\circ$}C}) at \SI{910}{MHz} and \SI{875}{MHz}, with switching activities extracted from post-\gls{PNR} gate-level simulations.

\subsection{Area Analysis and Breakdown}
\begin{figure}[ht]
    \centering
    \includegraphics[width=\linewidth]{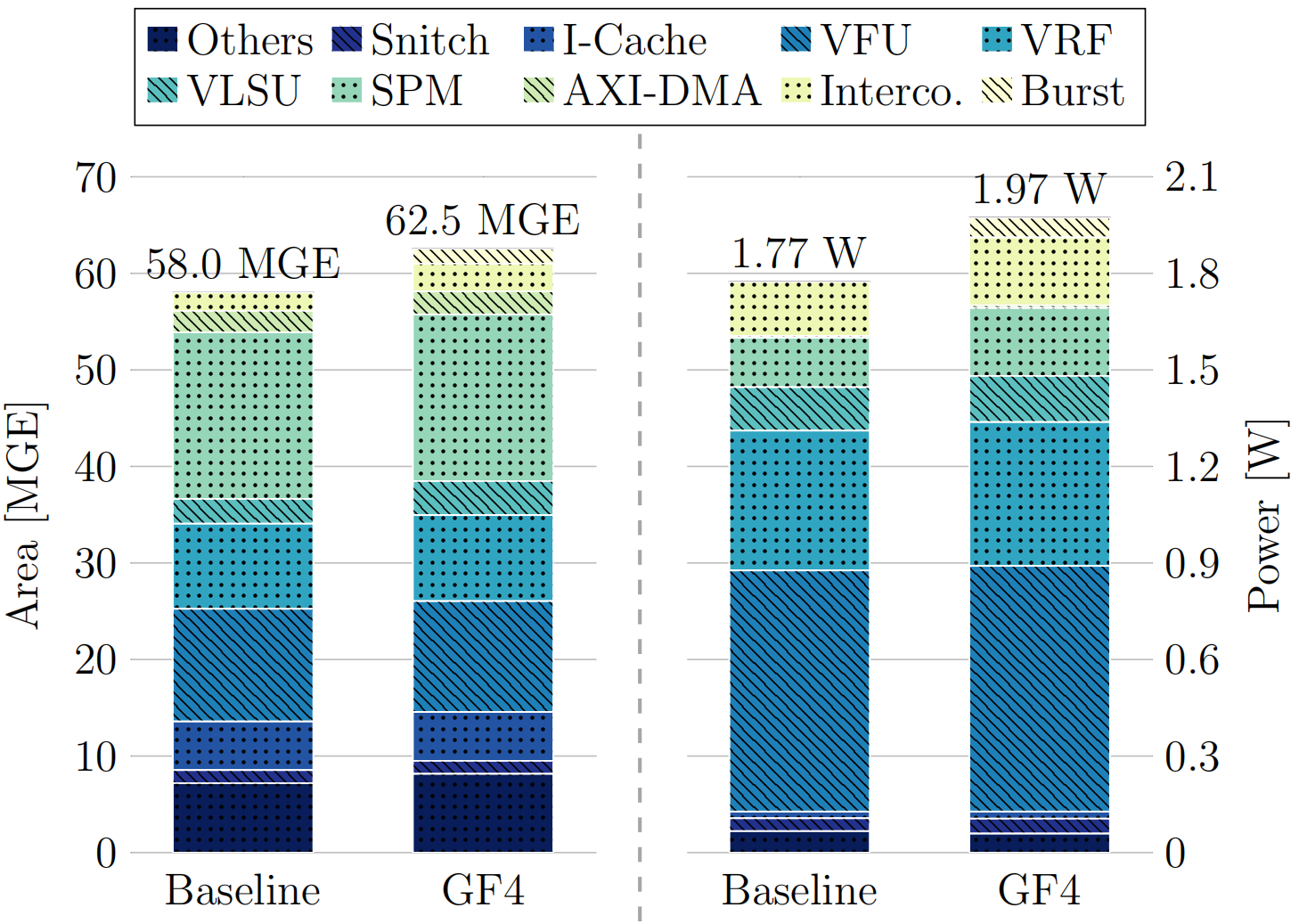}
    \caption{Area (left) and power (right) breakdown for the MemPool\textsubscript{64}Spatz\textsubscript{4} clusters. Area and power extracted in 12-nm technology, at $TT@910MHz$, executing \gls{MATMUL} kernel.}
    \label{fig:power-area}
\end{figure}

The post-\gls{PNR} physical layout of the GF4 design on MP\textsubscript{64}Spatz\textsubscript{4} is shown in~\Cref{fig:layout}.
Implementing the \gls{TCDM} Burst Access results in less than \SI{8}{\percent} logic area increase in all three clusters.
An area breakdown of MP\textsubscript{64}Spatz\textsubscript{4} with \textit{GF4} design is shown in the left part of~\Cref{fig:power-area}, with total area increased by \SI{4.5}{MGE}.
The \SI{35}{\percent} area increase in the \gls{VLSU} is primarily due to the enlarged \gls{ROB}. 
The increased data width in the response channel leads to a \SI{51}{\percent} logic area increase in the interconnection network.
The Burst Manager and the Burst Sender blocks contribute an additional \SI{1.5}{MGE} of the logic area in total, occupied mainly by the \glspl{FIFO} in the Burst Manager.

\subsection{Power Analysis and Breakdown}
We measure the power consumption per kernel in nominal operating conditions as recorded in~\Cref{tab_kernel} across different testbed clusters.
A power breakdown is shown in the right part of~\Cref{fig:power-area} on the GF4 design on MP\textsubscript{64}Spatz\textsubscript{4} testbed running $256\times 256\times 256$ \gls{MATMUL} kernel.
The increased power consumption in the \gls{VLSU}, \gls{SPM} banks and interconnection logic indicates a higher data transfer rate due to the increased hierarchical average bandwidth.
Even in this compute-bound kernel, which cannot benefit from higher bandwidth, we only observe a small reduction (less than \SI{8}{\percent} on average) in energy efficiency.

The memory-bound kernels show higher power consumption because of the higher \gls{FPU} utilization.
Compared to the baseline, a large energy efficiency gain, up to \SI{90}{\percent} improvement on a performance improvement of \SI{176}{\percent}, is observed in all kernels, for different scales of testbed cluster.


\section{Conclusion}
In this paper, we presented the \gls{TCDM} Burst Access, a software-transparent burst transaction architecture enhancement for bandwidth utilization improvement in many-core vector clusters with tightly coupled L1 memory.
By sending \SI{32}{b} narrow burst requests through the Burst Sender, adapting them to the \gls{SPM} banks via the Burst Manager, and adding a parametrizable datFa width on the response channel, \gls{TCDM} Burst Access significantly enhanced bandwidth utilization while maintaining scalability across different cluster scales.
We evaluated our design by implementing it into three sizes of the MemPool-Spatz architecture, validated in an advanced 12-nm technology node.
Our design improved the \gls{MATMUL} kernel performances up to \SI{62}{\percent}, fully pushing the kernels into the memory-bound region with less than a \SI{8}{\percent} increase in area.
Additionally, it achieved up to \textbf{2.76x} performance and \textbf{1.9x} energy efficiency improvements on memory-bound kernels compared to the baseline testbed clusters.

\section*{Acknowledgment}
    This work is funded in part by the COREnext project supported by the EU Horizon Europe research and innovation programme under grant agreement No. 101092598.

\bibliographystyle{IEEEtran}
\bibliography{source/reference}

\begin{thebibliography}{10}
\providecommand{\url}[1]{#1}
\csname url@samestyle\endcsname
\providecommand{\newblock}{\relax}
\providecommand{\bibinfo}[2]{#2}
\providecommand{\BIBentrySTDinterwordspacing}{\spaceskip=0pt\relax}
\providecommand{\BIBentryALTinterwordstretchfactor}{4}
\providecommand{\BIBentryALTinterwordspacing}{\spaceskip=\fontdimen2\font plus
\BIBentryALTinterwordstretchfactor\fontdimen3\font minus \fontdimen4\font\relax}
\providecommand{\BIBforeignlanguage}[2]{{%
\expandafter\ifx\csname l@#1\endcsname\relax
\typeout{** WARNING: IEEEtran.bst: No hyphenation pattern has been}%
\typeout{** loaded for the language `#1'. Using the pattern for}%
\typeout{** the default language instead.}%
\else
\language=\csname l@#1\endcsname
\fi
#2}}
\providecommand{\BIBdecl}{\relax}
\BIBdecl

\bibitem{Sevilla2022}
J.~Sevilla, L.~Heim, A.~Ho, T.~Besiroglu, M.~Hobbhahn, and P.~Villalobos, ``Compute trends across three eras of machine learning,'' in \emph{Proceedings of the International Joint Conference on Neural Networks}, vol. 2022-July, 2022.

\bibitem{llmsurvey2024}
\BIBentryALTinterwordspacing
S.~Minaee, T.~Mikolov, N.~Nikzad, M.~Chenaghlu, R.~Socher, X.~Amatriain, and J.~Gao, ``Large language models: A survey,'' 2024. [Online]. Available: \url{https://arxiv.org/abs/2402.06196}
\BIBentrySTDinterwordspacing

\bibitem{Tenstorrent}
J.~Vasiljevic and D.~Capalija, ``Blackhole \& tt-metalium: The standalone ai computer and its programming model,'' in \emph{2024 IEEE Hot Chips 36 Symposium (HCS)}, 2024, pp. 1--30.

\bibitem{ETSOC}
D.~R. Ditzel and the Esperanto~team, ``Accelerating ml recommendation with over 1,000 risc-v/tensor processors on esperanto's et-soc-1 chip,'' \emph{IEEE Micro}, vol.~42, no.~3, pp. 31--38, 2022.

\bibitem{A64FX}
S.~Matsuoka, ``Fugaku and a64fx: the first exascale supercomputer and its innovative arm cpu,'' in \emph{2021 Symposium on VLSI Circuits}, 2021, pp. 1--3.

\bibitem{Kalray}
\BIBentryALTinterwordspacing
B.~D. de~Dinechin, ``Consolidating high-integrity, high-performance, and cyber-security functions on a manycore processor,'' in \emph{Proceedings of the 56th Annual Design Automation Conference 2019}, ser. DAC '19.\hskip 1em plus 0.5em minus 0.4em\relax New York, NY, USA: Association for Computing Machinery, 2019. [Online]. Available: \url{https://doi.org/10.1145/3316781.3323473}
\BIBentrySTDinterwordspacing

\bibitem{terapool}
\BIBentryALTinterwordspacing
Y.~Zhang, M.~Bertuletti, S.~Riedel, M.~Cavalcante, A.~Vanelli-Coralli, and L.~Benini, ``Terapool-sdr: An 1.89tops 1024 rv-cores 4mib shared-l1 cluster for next-generation open-source software-defined radios,'' in \emph{Proceedings of the Great Lakes Symposium on VLSI 2024}.\hskip 1em plus 0.5em minus 0.4em\relax ACM, 6 2024, pp. 86--91. [Online]. Available: \url{https://dl.acm.org/doi/10.1145/3649476.3658735}
\BIBentrySTDinterwordspacing

\bibitem{mempool}
S.~Riedel, M.~Cavalcante, R.~Andri, and L.~Benini, ``Mempool: A scalable manycore architecture with a low-latency shared l1 memory,'' \emph{IEEE Transactions on Computers}, vol.~72, 2023.

\bibitem{task_schedule_2017}
\BIBentryALTinterwordspacing
T.~Kim, J.~Lim, J.~Kim, W.-C. Cho, E.-Y. Chung, and H.-J. Lee, ``Scalable bandwidth shaping scheme via adaptively managed parallel heaps in manycore-based network processors,'' \emph{ACM Trans. Des. Autom. Electron. Syst.}, vol.~22, no.~4, 2017. [Online]. Available: \url{https://doi.org/10.1145/3065926}
\BIBentrySTDinterwordspacing

\bibitem{task_schedule_2018}
Y.~Hu and T.~Li, ``Enabling efficient network service function chain deployment on heterogeneous server platform,'' in \emph{2018 IEEE International Symposium on High Performance Computer Architecture (HPCA)}, 2018, pp. 27--39.

\bibitem{data_arrange_2023}
J.~Wang, Z.~Wang, and Y.~Hu, ``Towards an efficient simd virtual radio access network (vran) and edge cloud system,'' \emph{IEEE Transactions on Cloud Computing}, vol.~11, no.~3, pp. 3226--3238, 2023.

\bibitem{FlooNoC_2023}
T.~Fischer, M.~Rogenmoser, M.~Cavalcante, F.~K. Gürkaynak, and L.~Benini, ``Floonoc: A multi-tb/s wide noc for heterogeneous axi4 traffic,'' \emph{IEEE Design and Test}, vol.~40, no.~6, pp. 7--17, 2023.

\bibitem{spatz}
M.~Cavalcante, D.~Wüthrich, M.~Perotti, S.~Riedel, and L.~Benini, ``Spatz: A compact vector processing unit for high-performance and energy-efficient shared-l1 clusters,'' in \emph{IEEE/ACM International Conference on Computer-Aided Design, Digest of Technical Papers, ICCAD}.\hskip 1em plus 0.5em minus 0.4em\relax Institute of Electrical and Electronics Engineers Inc., 10 2022.

\bibitem{roofline}
\BIBentryALTinterwordspacing
S.~Williams, A.~Waterman, and D.~Patterson, ``Roofline: an insightful visual performance model for multicore architectures,'' \emph{Commun. ACM}, vol.~52, no.~4, p. 65–76, apr 2009. [Online]. Available: \url{https://doi.org/10.1145/1498765.1498785}
\BIBentrySTDinterwordspacing

\end{thebibliography}

\end{document}